\documentclass[a4paper,11pt]{article}
\usepackage{amssymb,amsmath,bm}
\usepackage{graphics,graphicx}
\usepackage{array,booktabs}
\usepackage{authblk}
\usepackage{slashed}
\usepackage{subfig}
\usepackage{cite}
\usepackage[margin=2.5cm]{geometry}
\usepackage[table,dvipsnames]{xcolor} 
\definecolor{Dblue}{rgb}{0.1, 0.1, 0.8}
\definecolor{Lblue}{rgb}{0.22,0.51,0.9}
\definecolor{green}{RGB}{53,170,102}
\definecolor{Green}{rgb}{0.0, 0.5, 0.0}
\definecolor{Bred}{rgb}{0.8, 0.25, 0.33}
\usepackage{tikz}
\usepackage{multirow}
\usepackage{tikz-feynman}
\usepackage[linktocpage,
colorlinks = true,
linkcolor = Bred,
urlcolor  = Dblue,
citecolor = green,
anchorcolor = blue]{hyperref}
\usepackage{epsfig}
\numberwithin{equation}{section}
\usepackage{multirow}
\usepackage{cite}
\usepackage{xcolor}
\usepackage[margin=2.5cm]{geometry}
\usepackage{orcidlink}
\usepackage[utf8]{inputenc}
\usepackage{orcidlink}

\DeclareUnicodeCharacter{2212}{-}

\numberwithin{equation}{section}

\usepackage{epsfig}

\title{\textbf{The vacuum stability and the hierarchy problem in a fermionic dark matter model}}

\author[a]{ Mojtaba Hosseini\thanks{\href{mailto:mojtaba\textunderscore hosseini@semnan.ac.ir}{mojtaba\textunderscore hosseini@semnan.ac.ir}}}

\affil[a]{Department of Physics, Semnan University, P.O. Box 35131-19111, Semnan, Iran}
\date{\today}

\begin{document}

\baselineskip 0.6 cm
\maketitle

\begin{abstract}
We consider an extension to the Standard Model (SM) with four new fields including scalar($S$), spinor($\psi^{1,2}$) and vector($V_\mu$)
under new $U(1)$ gauge group in the hidden sector. The scalar particle interacts with the SM Higgs particle and is an intermediary between the dark and the SM parts .
Our dark matter(DM) candidate is the spinor particle.
We show that the model successfully explains the relic density of the DM in the universe and evades
the strong bounds from direct detection experiments while respecting the theoretical constraints and the vacuum stability conditions.
In addition, we study the hierarchy problem within the Veltman approach by solving the renormalization group equations at one-loop.
We demonstrate that the addition of the new fields contributes to the Veltman parameters which in
turn results in satisfying the Veltman conditions much lower than the Planck scale.
For the our DM model we find one representative point in the viable parameter space which satisfy also the Veltman conditions at $\Lambda$ = 1 TeV.
Therefore, the presence of the extra particle solves the fine-tuning problem of the Higgs mass.
\end{abstract}



\section{Introduction} \label{sec1}
The Standard Model of elementary particles(SM) currently provides the best description of the fundamental particles and interactions of the universe.
In fact, the discovery of the Higgs boson at the Large Hadron Collider (LHC) completed the last remaining piece of the SM and established the SM as a renormalizable theory\cite{ATLAS:2012yve,CMS:2012qbp,CMS:2020xrn,ATLAS:2018tdk}.
However, this model will not be considered a final model because it has shortcomings in both the theoretical and experimental parts.
For this reason, models beyond the standard model(BSM) are proposed and investigated to address these deficiencies in both of the aforementioned parts.

One of the issues in the theoretical part that poses a problem for the SM is the instability of the Higgs potential at high energies.
In the SM, the Higgs quartic coupling becomes negative at the scale $10^{10}$ GeV, and the Higgs non-zero vacuum expectation value(VEV) is no longer a minimum of the theory. The reason for this is that the top quark has a large negative contribution to the Renormalization Group Equation (RGE) for $\lambda_{H}$.
Therefore, extensions to the SM by adding new particles such as scalars could be a solution to this issue\cite{Gonderinger:2012rd,Abada:2013pca,Baek:2012uj,YaserAyazi:2022tbn,Ghorbani:2017qwf}.
An important question to pose is up to what scale we should expect a model to be valid
and the common investigation for vacuum stability analyses in the literature begins with the
RGE improved potential and choice of the renormalization scale to minimize the one-loop potential.

Another problem in the theoretical part of the SM is the hierarchy problem.
In the SM, the Higgs mass receives additive quantum corrections proportional to the momentum cut-off squared, $\Lambda^2$
at one-loop level. This quadratically divergent one-loop correction to the Higgs mass was first calculated by Veltman\cite{Veltman:1980mj}.
If $\Lambda$ is very large, the Higgs mass suffers from the quadratic divergence of this correction, which addresses
the naturalness issue of the theory resulting in the fine-tuning problem\cite{Susskind:1978ms}.
Veltman has suggested the cancellation of the quadratic divergence by itself leading to the prediction
of the Higgs mass, which is called the Veltman condition (VC).
In fact, Veltman found that the Higgs mass should be $\sim$ 314 GeV, which is of course not consistent with the observed value.
In order to impose the VC to the Higgs mass, required are some new degrees of freedom which interact with the Higgs boson to contribute to
the Higgs self-energy diagram \cite{Gunion:2008kp}.  Simple extensions of the SM resolving the fine-tuning problem by applying the VC
have been studied widely in many literature\cite{Masina:2013wja,ElHedri:2018hix,Katz:2005au,Kim:2018ecv,Chakraborty:2012rb,aali:2020tgr,Bian:2013xra,Karahan:2014ola,Chakrabarty:2018yoy,Darvishi:2017bhf,Chakraborty:2014xqa,Bazzocchi:2012pp,Ma:2014cfa,Abu-Ajamieh:2021vnh,Das:2023zby}.
In practice, it is assumed that the SM is an effective field theory valid up to a scale $\Lambda$, with $\Lambda$ being the scale of
new physics beyond the SM. Then the Higgs mass radiative corrections of $\mathcal{O}$($\Lambda^2$), for the leading contributions, receives a huge fine-tuning mechanism to keep the renormalized mass as small as the measured Higgs mass.
Of course, A prominent mechanism to protect the Higgs mass correction
from large amounts is supersymmetric scenarios that unfortunately, no clues have been found in the experiments so far.
Another well known solution to the fine-tuning problem is the scale invariant models\cite{YaserAyazi:2024dxk,Hosseini:2024ahe,Hosseini:2023qwu,YaserAyazi:2019caf,YaserAyazi:2018lrv,Abkenar:2024ket,Plascencia:2015xwa,Kobakhidze:2014afa,Antipin:2013exa,Ghorbani:2015xvz,Ghorbani:2022muk,Habibolahi:2022rcd}.

One of the most important shortcomings of the SM is the lack of a particle that can account for the amount of DM observed in the universe.
 DM is estimated to make up approximately 27 percent of our universe, as indicated by a lot of astrophysical and cosmological evidence\cite{Bertone:2016nfn}.
Weakly interacting massive particles (WIMPs) are the most popular candidate for DM, with the Freeze-out scenario being the most popular choice\cite{Feng:2022rxt}.
WIMP paradigm is essential background for almost any discussion of particle DM and the triple coincidence of motivations from particle theory, particle experiment, and cosmology is known as the WIMP miracle.
In the present work we exploit a DM model with three additional fields vector, scalar and spinor that fermionic particle is DM candidate and the scalar particle is the intermediary between the SM and the DM. By running the couplings at high scale we study the hierarchy problem and the effect of extra
fields on the Veltman conditions while respecting the stability of the extended Higgs potential.
As experimental bounds, we consider strong constraints from observed relic
density granted by PLANCK observations, bounds from direct detection experiments like XENONnT  and the invisible Higgs decay width.
On the other hand, theoretical constraints such as the perturbativity and positivity condition are also examined.

The setup of the paper is the following. In the next section we describe the fermionic DM model and impose theoretical constraints.
Section \ref{DM phenomenology} is devoted to the phenomenology of the DM including bounds from the observed relic density, direct detection experiments and
the invisible Higgs decay and the permissible parameter space of the model is shown.
The RGEs and the numerical results for running of the couplings are given in section \ref{RGEs} and the vacuum stability of the potential of the model is investigated.
The relevant Veltman condition for the SM Higgs and the intermediate scalar
are obtained in section \ref{Veltman condition}. We conclude our results in section \ref{Conclusion}.

\section{The model}\label{Model}
We propose a model comprising two spinor fields $\psi^{1,2}$ as DM(including
right-handed $\psi^{1,2}_R$ and  left-handed $\psi^{1,2}_L$), and a vector field $V_\mu$, along with a
(complex) scalar field $S$ serving as an intermediate particle. All of these new fields are singlet under SM gauge group.
We implement a discrete symmetry under which, the new fields transform as follows
\begin{equation}\label{symmetry}
S\longrightarrow S^*,~~~  V_\mu \longrightarrow V_\mu,~~~ \psi^1_L \longrightarrow -\psi^2_L,~~~~ \psi^1_R \longrightarrow -\psi^2_R,
\end{equation}
while all SM particles are even. The scalar field $S$ and spinor fields ($\psi^{1,2}_R$ and $\psi^{1,2}_L$) carry charge under a dark $U_D(1)$ gauge symmetry,
 with the vector field $V_\mu$ serving as the gauge field.  All SM particles are singlets under the dark
gauge symmetry. The dark sector is invariant under the transformations of $U_D(1)$ gauge group:
\begin{align} \label{2-2}
& \psi^a_L \rightarrow e^{i Q^a_L \alpha(x)} \psi^a_L, \nonumber \\
& \psi^a_R \rightarrow e^{i Q^a_R \alpha(x)} \psi^a_R, \nonumber \\
& S \rightarrow e^{i Q_S \alpha(x)} S, \nonumber \\
& V_{\mu} \rightarrow V_{\mu} - \frac{1}{g_v} \partial_{\mu}{\alpha(x)},
\end{align}
where the $U(1)_D$ charge of the new particles, $ Q_{L,R,S} $ are given in Table \ref{Table} and $a\in \{1,2\}$.

\begin{table}
\begin{center}
\begin{tabular}{| l | l | l | l | l | l | l |}
\hline
Field&$S$&$V_\mu$&$\psi^1_L$&$\psi^1_R$&$\psi^2_L$&$\psi^2_R$\\ \hline
$U(1)_D$ charge($Q$)&1&0&$\frac{1}{2}$&$-\frac{1}{2}$&$-\frac{1}{2}$&$\frac{1}{2}$\\ \hline
\end{tabular}
\end{center}
\caption{\label{Table}The charges of the dark sector particles under the new $U(1)_D$ symmetry.}
\end{table}

The most general Lagrangian by renormalizable interactions is given by:
\begin{align}
&{\cal L} ={\cal L}_{SM}-\frac{1}{4} V_{\mu \nu} V^{\mu \nu}+ (D_{\mu} S)^{*} (D^{\mu} S)- V(H,S) \nonumber \\+
& \sum_{a=1}^{2}(i\bar\psi^a_L  \gamma^{\mu}D_{\mu}\psi^a_L+ i\bar\psi^a_R  \gamma^{\mu}D_{\mu}\psi^a_R)
-g_{s,1} S\bar\psi^1_L \psi^1_R -g_{s,2} S^* \bar\psi^2_L \psi^2_R+H.C.,
\label{eq:lagrangian}
\end{align}
where $ {\cal L} _{SM} $ is the SM Lagrangian without the Higgs potential term. The covariant derivative is
\begin{align}
& D_{\mu}= (\partial_{\mu} + i Q g_{v} V_{\mu}), \quad \text{and}\nonumber \\
& V_{\mu \nu}= \partial_{\mu} V_{\nu} - \partial_{\nu} V_{\mu}.\end{align}
The lagrangian presented \ref{eq:lagrangian} is invariant under both discrete symmetry \ref{symmetry} and local gauge transformation \ref{2-2}.
Finally, the most general scale-invariant potential $ V(H,S) $ which is renormalizable and invariant
under gauge symmetry is
 \begin{equation}
V(H,S) = -\mu_{H}^2 H^{\dagger}H-\mu_{S}^2 S^*S+\lambda_{H} (H^{\dagger}H)^{2} + \lambda_{S} (S^*S)^{2} +  \lambda_{S H} (S^*S) (H^{\dagger}H). \label{2-5}
\end{equation}
 SM Higgs field $ H $, as well as dark scalar $S$, can receive VEVs breaking respectively the electroweak and $ U_{D}(1) $ symmetries.
In the unitary gauge, the imaginary component of $S$ can be absorbed as the longitudinal component of $ V_{\mu} $.
In this gauge, we can write
\begin{equation}
H = \frac{1}{\sqrt{2}} \begin{pmatrix}
0 \\ h_{1} \end{pmatrix} \, \, \, {\rm and} \, \, \, S = \frac{1}{\sqrt{2}} h_{2} , \label{2-6}
\end{equation}
where $ h_{1} $ and $ h_{2} $ are real scalar fields which can get VEVs. The tree level potential in unitary gauge is:
\begin{equation}
V_{\text{tree}}(h_{1},h_{2})=-\frac{1}{2} \mu _H^2 h_1^2-\frac{1}{2} \mu _S^2 h_2^2 +\frac{1}{4} \lambda _H h_1^4 +\frac{1}{4} \lambda _S h_2^4+\frac{1}{4} \lambda _{SH} h_1^2 h_2^2.
\end{equation}
In order to get the mass spectrum of the model, it is necessary to consider the sufficient conditions for a local minimum:
\begin{align}
& \quad \dfrac{\partial V_{\text{tree}}}{\partial h_1}\bigg|_{h_1=h_2=0}=0 ,\\
& \quad \dfrac{\partial V_{\text{tree}}}{\partial h_2}\bigg|_{h_1=h_2=0}=0 \label{minimum1} ,\\
&  \det {\cal{H}} > 0 \label{minimum2} \\
& {\cal{H}}_{11} > 0 \label{minimum3}
\end{align}
to occur at a point $ (\nu_{1},\nu_{2}) $. In the equations above $ {\cal{H}}_{ij}(h_{1},h_{2})=\frac{\partial^{2}V_{\text{tree}}}{\partial h_{i} \partial h_{j} }$ is the Hessian matrix.  Note that Eq. (\ref{minimum2}) and Eq. (\ref{minimum3}) also imply that $ {\cal{H}}_{22} > 0 $.
Eq. (\ref{minimum1}) leads to
\begin{align}
& \mu _H^2= \lambda _H \nu _1^2+ \frac{1}{2} \lambda _{SH} \nu _2^2 , \nonumber\\
& \mu _S^2=\lambda _S \nu _2^2+ \frac{1}{2} \lambda _{SH} \nu _1^2 . \label{mini}
\end{align}
Eq. (\ref{mini}) leads to the non-diagonal mass matrix $\cal{H}$ as follows:
\begin{equation}
{\cal{H}}(\nu_{1},\nu_{2})= \left(
\begin{array}{cc}
 2 \lambda _H \nu _1^2 & \lambda _{SH} \nu _1 \nu _2 \\
 \lambda _{SH} \nu _1 \nu _2 & 2 \lambda _S \nu _2^2 \\
\end{array}
\right). \label{hess}
\end{equation}
Now by substituting $ h_1 \rightarrow \nu_1 + h_1 $ and $ h_2 \rightarrow \nu_2 + h_2 $, the fields $ h_1 $ and $ h_1 $ mix with each other and they can be rewritten by the mass eigenstates $ H_1 $ and $ H_1 $ as
\begin{equation}
\begin{pmatrix}
h_{1}\\h_{2}\end{pmatrix}
 =\begin{pmatrix} cos \alpha~~~  sin \alpha \\-sin \alpha  ~~~~~cos \alpha
 \end{pmatrix}\begin{pmatrix}
H_1 \\  H_{2}
\end{pmatrix}, \label{matri}
\end{equation}
where $ \alpha $ is the mixing angle. After symmetry breaking, we have
\begin{align}
& \nu_{2} =  \frac{M_{V}}{g_v} , &\nonumber
& sin \alpha =  \frac{\nu_{1}}{\sqrt{\nu_{1}^{2}+\nu_{2}^{2}}} \nonumber \\
&M_{\psi}= \frac{g_sM_{V}}{\sqrt{2}g_v}&\nonumber
& \lambda _H=\frac{\cos ^2\alpha M_{H_1}^2+\sin ^2\alpha  M_{H_2}^2}{2 \nu _1^2}  \nonumber  \\
& \lambda _S=\frac{\sin ^2\alpha M_{H_1}^2+\cos ^2\alpha  M_{H_2}^2}{2 \nu _2^2} &\nonumber
&  \lambda _{SH}=\frac{ \left(M_{H_2}^2-M_{H_1}^2\right) \sin \alpha  \cos \alpha}{\nu _1 \nu _2}, \\ \label{2-15}
\end{align}
where $M_{\psi}$ and $M_{V}$ are the masses of vector and fermion fields after symmetry breaking. Of course, Due
to the symmetry condition (\ref{symmetry}), it follows that $g_{s,1} = g_{s,2}=g_s$, which consequently
implies $M_{\psi_1}=M_{\psi_2}=M_\psi$. It is necessary to mention this point as mentioned in reference \cite{Abkenar:2024ket} that
in our model, the dark $U_D(1)$ symmetry introduces a new gauge group, necessitating the examination of the cubic anomaly $[U_D(1)]^3$
from  $U_D(1)$ -charged fermions. In the presented model, the $[U_D(1)]^3$ anomaly is canceled by the presence of two fermions.

There are some theoretical constraints on the model structure and its stability. The  perturbativity condition of the couplings, includes the following relations:
\begin{align}
&-4\pi < \lambda_{SH} ,~g_s ,~g_v <4\pi  \nonumber \\
&0<  \lambda_{H} ,~\lambda_{S}< 4\pi. \label{2-16}
\end{align}
We also require that the vacuum be stable (bounded from below) at tree level which implies
\begin{equation}
\lambda_H>0   ,  \lambda_S>0  , \lambda_{SH}>-2 \sqrt{\lambda_H \lambda_S }. \label{2-17}
\end{equation}
Note that according to (\ref{2-15}), the model has only four free parameters $g_v$, $M_\psi$, $M_{H_2}$, and $M_V$.
In the following, we examine the phenomenology of model including DM relic
density and data from direct detection experiments.

\section{DM phenomenology}\label{DM phenomenology}
\subsection{Relic density}
\begin{figure}
	\begin{center}
		\centerline{\hspace{0cm}\epsfig{figure=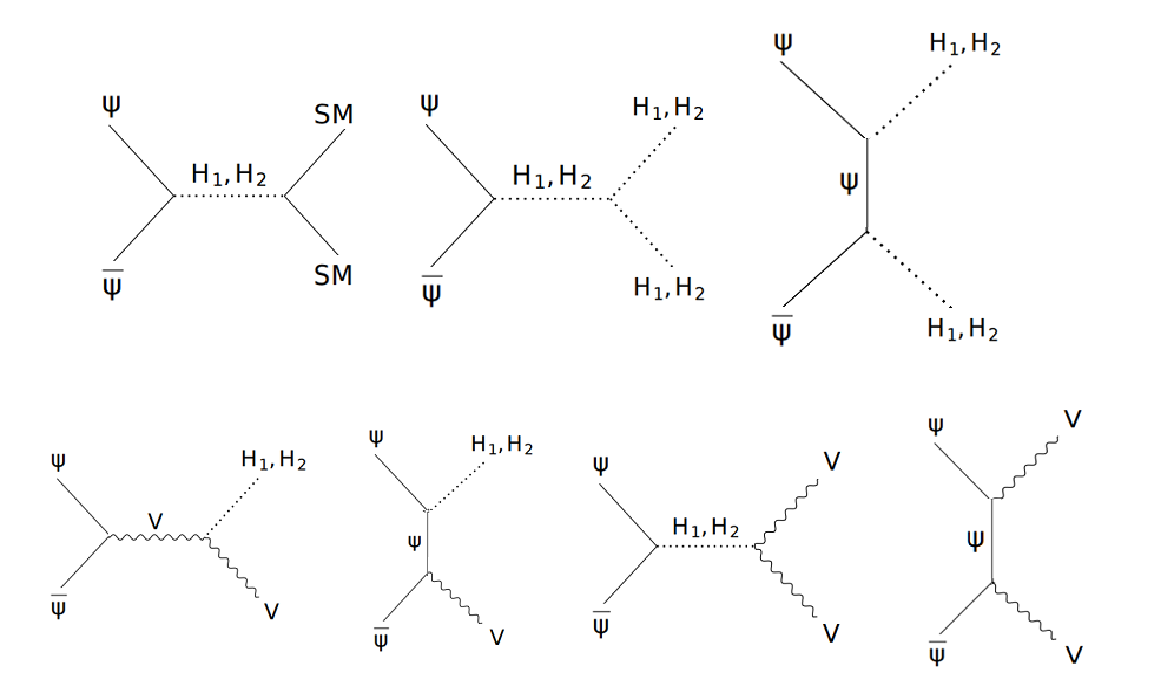,width=18cm}}
        \centerline{\vspace{-0.1cm}}
		\caption{The relevant Feynman diagrams for dark matter annihilation at the freeze-out epoch.} \label{feynman}
	\end{center}
\end{figure}

\begin{figure}
	\begin{center}
		\centerline{\hspace{0cm}\epsfig{figure=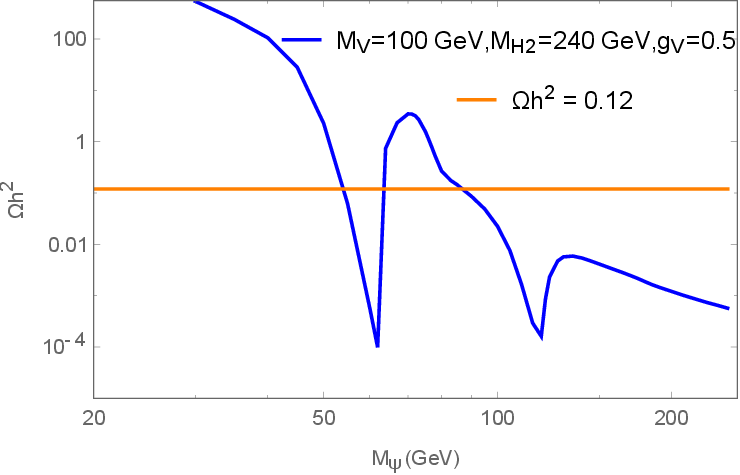,width=12cm}}
        \centerline{\vspace{-0.1cm}}
		\caption{Relic density of DM as a function of the DM mass.} \label{omegafermion}
	\end{center}
\end{figure}

 It is assumed that the DM candidate is thermal relic, so first early Universe is hot and very dense and all particles are in thermal equilibrium.
 As the universe expands, the DM particles are diluted and can no longer find each other until they are annihilated and are out of thermal equilibrium with the SM particles.
 Then the DM particles freeze out and their number asymptotically reaches a constant value as their thermal relic density.
 DM number density, $(n_{\psi})$, is governed by the Boltzmann equation:
\begin{equation} \label{3-1}
\dot{n_{\psi}} + 3Hn_{\psi} = -\langle\sigma_{ann} \nu_{rel}\rangle [n_{\psi} ^2 -(n_{\psi} ^{eq})^2] ,
\end{equation}
where $H$ is the Hubble parameter and $n_{\psi} ^{eq} \sim (m_{\psi} T)^{3/2} e^{-m_{\psi} /T} $ is the particle density before particles get out of equilibrium.
In our model, the DM fermions $\psi^{1,2}$ possess identical masses and
contribute equally to the relic abundance of fermionic DM ($\psi$).
The relevant Feynman diagrams for computation of the DM annihilation cross section is depicted in Fig\ref{feynman}.
In figure \ref{omegafermion} we have plotted DM relic density diagram versus the DM mass. As can be seen there are two minima
observed in the DM relic density around 62 GeV and 120 GeV. These reductions in relic density at these points are associated with the resonances
of $H_1$ and $H_2$, which is a general feature of purely scalar-portal DM scenarios that $\Omega h^2 \sim 0.1$(observed relil density) majorly occurs near the
resonance dips.

\subsection{Direct detection}
\begin{figure}
	\begin{center}
		\centerline{\hspace{0cm}\epsfig{figure=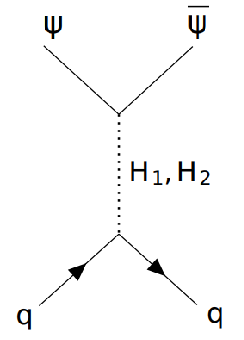,width=5cm}}
        \centerline{\vspace{-0.1cm}}
		\caption{ Relevant Feynman diagram for direct detection.} \label{direct}
	\end{center}
\end{figure}

In our model, the DM particles interact with quarks via the exchange of $H_1$ or
$H_2$ within the Higgs portal framework, leading to a spin-independent cross section for DM-nucleon interactions(see figure \ref{direct}).
The DM elastic scattering cross section is spin-independent (SI) and is given by
\begin{equation}
\sigma_{\psi-N} ^{SI} =\frac{g_s^2 \sin ^2\alpha \cos ^2\alpha}{2 \nu_1^2}(\frac{1}{M_{H1} ^2}-\frac{1}{M_{H2} ^2})^2 \frac{M_N ^4 M_{\psi}^2}{\pi (M_N +M_\psi)^2}f_N ^2 ,
\end{equation}
where $M_N$ represents the mass of the nucleon and $f_N$ = 0.3 characterizes the coupling
between the Higgs and nucleons.

\subsection{Invisible Higgs decay}
In the model, SM Higgs-like($H_1$) can decay invisibly into a pair of DM, $VV$ and $H_2 H_2$ if kinematically allowed and be placed in the permissible parameter space.
The invisible Higgs decay width for decay a pair of DM is :
\begin{eqnarray}
\Gamma(H_1\rightarrow {\psi}{\psi})& =\frac{M_{H_1} g_s^2 \sin ^2\alpha}{8\pi} (1-\frac{4M_{{\psi}}^2}{{M_{H_1}^2}})^{3/2} .
\label{decay1}
\end{eqnarray}
The invisible Higgs decay width for the process $H_1 \rightarrow V V$ is
\begin{eqnarray}
\Gamma(H_1\rightarrow V V) =\frac{g_v^2 M_{H_1}^3 \sin ^2\alpha}{16\pi M_V^2}\sqrt{1-4(M_V / M_{H_1})^2} [1-4(M_V / M_{H_1})^2 +12(M_V / M_{H_1})^4].
\label{decay2}
\end{eqnarray}
The invisible Higgs decay width for the process $H_1 \rightarrow H_2 H_2$ is
\begin{eqnarray}
\Gamma(H_1\rightarrow H_2 H_2) =\frac{c^2}{8\pi M_{H_1}}\sqrt{1-4(M_{H_2} / M_{H_1})^2}.
\label{decay3}
\end{eqnarray}
where the coupling $c$ is given by the relation below
\begin{align}
& c= \lambda_{SH}v_1 \cos\alpha + 2\lambda_{SH}v_2 \sin\alpha -3v_2 \lambda_{SH} \sin ^3\alpha +6\lambda_H v_1  \sin ^2\alpha \cos\alpha \nonumber \\
&-3\lambda_{SH}v_1 \sin ^2\alpha \cos\alpha -6v_2 \lambda_S \cos ^2\alpha \sin\alpha.
\label{decay4}
\end{align}
Therefore, $H_1$ can contribute to the invisible decay mode with a branching ratio:
\begin{eqnarray}
Br(H_1\rightarrow \rm Invisible)& =\frac{\Gamma(H_1\rightarrow V V)+\Gamma(H_1\rightarrow H_2 H_2)+\Gamma(H_1\rightarrow \psi \psi)}{\Gamma(h)_{SM}+\Gamma(H_1\rightarrow V V)+\Gamma(H_1\rightarrow H_2 H_2)+\Gamma(H_1\rightarrow \psi \psi)},
\label{decay5}
\end{eqnarray}
where $\Gamma(h)_{SM}=4.15 ~ \rm [MeV]$ is total width of Higgs boson\cite{LHCHiggsCrossSectionWorkingGroup:2011wcg}.
CMS Collaboration has  reported the observed (expected) upper limit
on the invisible branching fraction of the Higgs boson to be $0.18 (0.10)$ at the $95\%$ confidence level, by assuming the SM production cross section \cite{CMS:2022qva}. A Similar analysis was performed by ATLAS collaboration in which an observed upper limit of $0.145$ is placed on the branching fraction of its decay into invisible particles at a $95\%$ confidence level\cite{ATLAS:2022yvh}. We use the results of the ATLAS experiment to constrain the parameter space.

We also used constraints related to mixing between scalars ($sin \alpha$) in checking our results. Non-observation of DM at the LHC corresponds to an upper bound on the value of $sin \alpha$. The Higgs boson data from LHC at 7 and 8 TeV sets a constraint of $|sin \alpha| < 0.36$ at 95\% C.L., independently of the
$H_2$ mass\cite{Carena:2018vpt}. There is also a mass-dependent constraint,
which requires $sin \alpha < 0.32$ for $m_{H2}> 200 GeV$ and $sin \alpha < 0.2$ for $m_{H2}> 400 GeV$, mostly
coming from restrictions on the NLO corrections to the mass of the W boson\cite{Robens:2015gla}.

\subsection{Results}
\begin{figure}
	\begin{center}
		\centerline{\hspace{0cm}\epsfig{figure=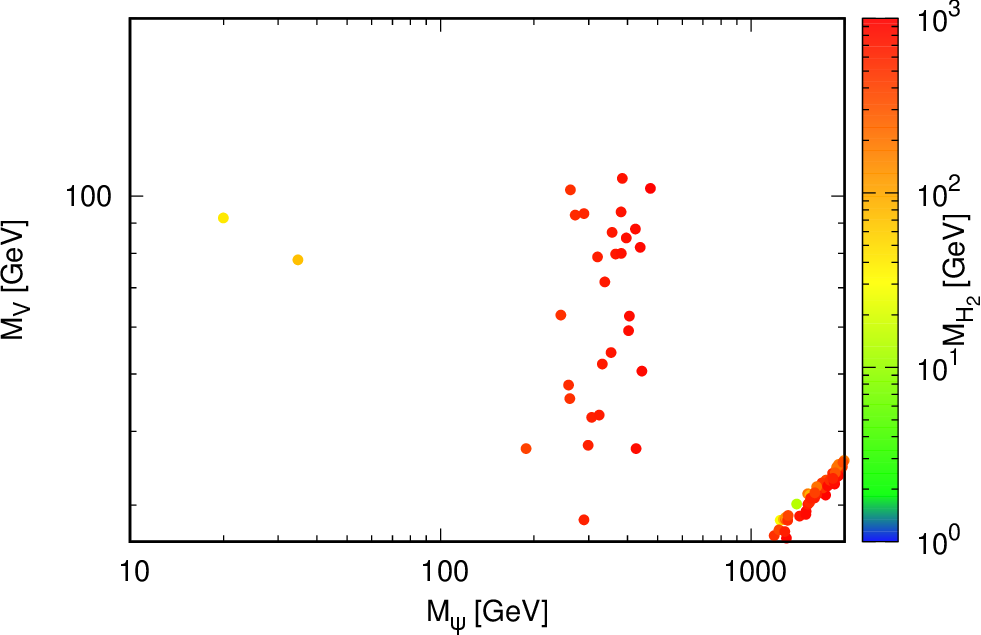,width=8cm}\hspace{0.5cm}\hspace{0cm}\epsfig{figure=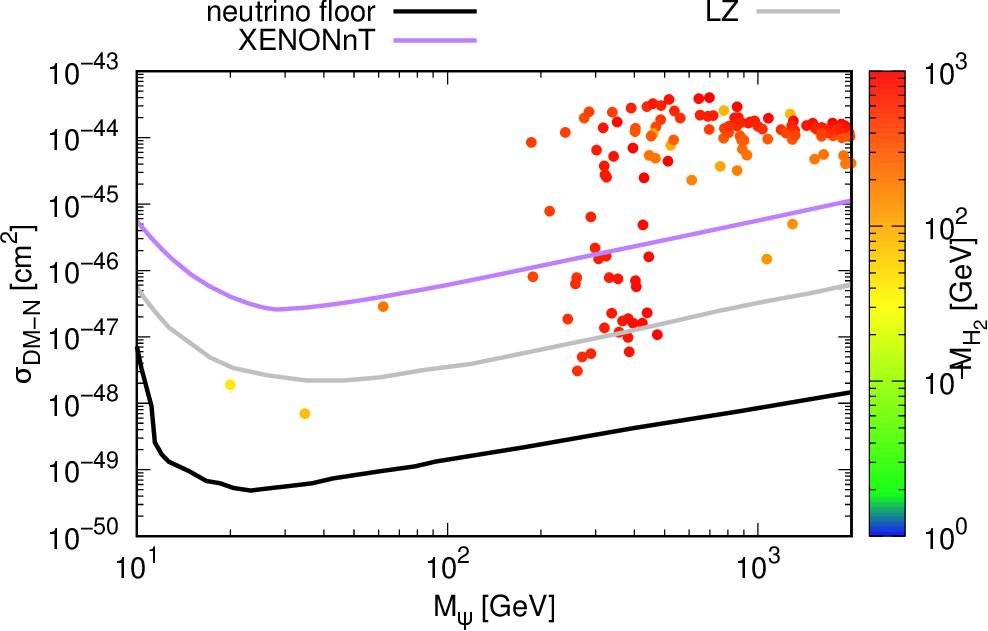,width=8cm}}
		\centerline{\vspace{0.2cm}\hspace{0.6cm}(a)\hspace{8cm}(b)} \centerline{\hspace{0cm}\epsfig{figure=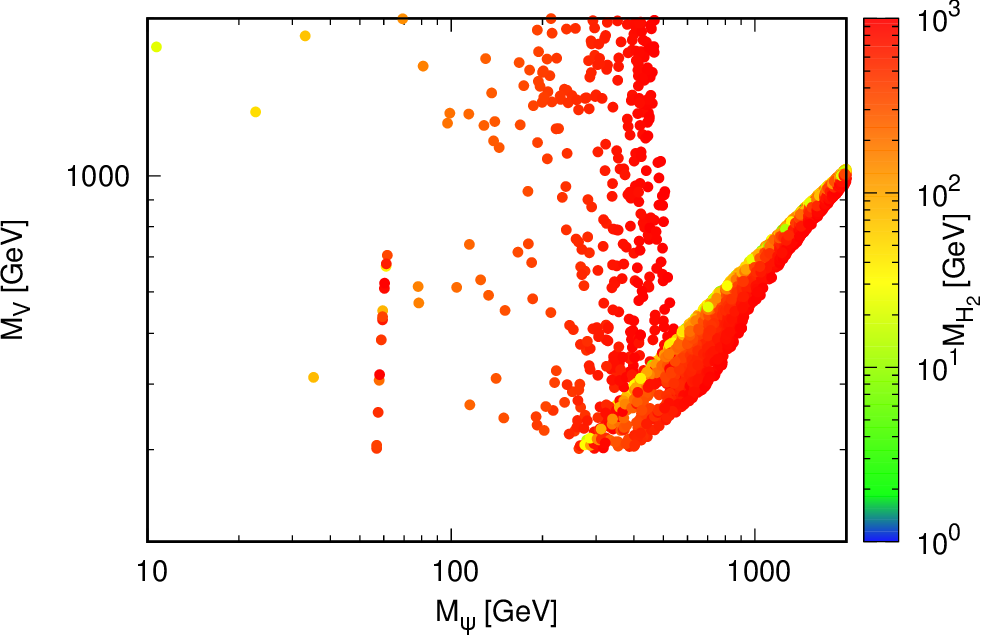,width=8cm}\hspace{0.5cm}\hspace{0cm}\epsfig{figure=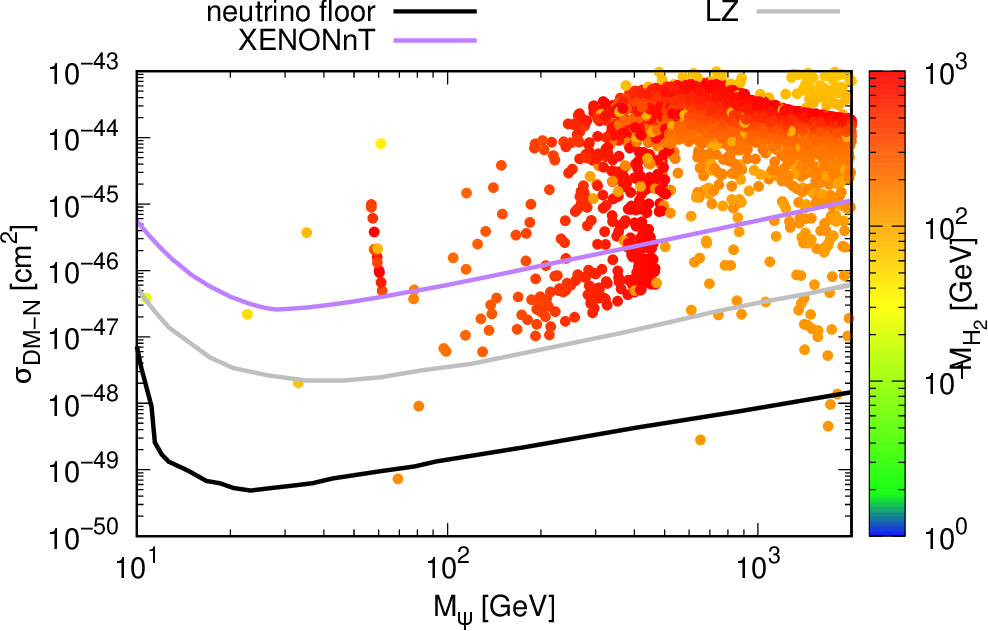,width=8cm}}		
		\centerline{\vspace{0.2cm}\hspace{0.6cm}(c)\hspace{8cm}(d)}
        \centerline{\hspace{0cm}\epsfig{figure=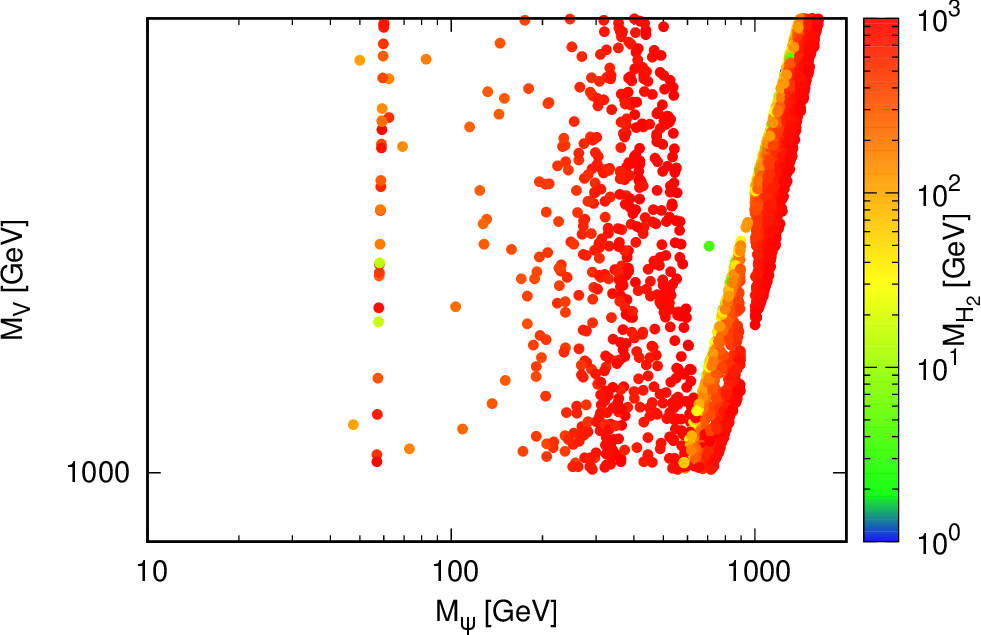,width=8cm}\hspace{0.5cm}\hspace{0cm}\epsfig{figure=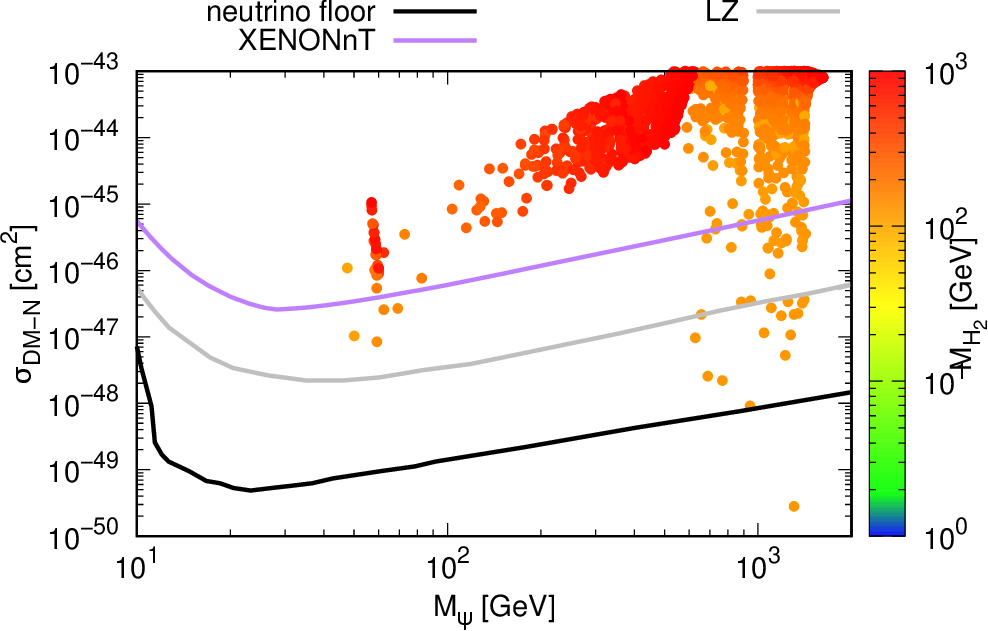,width=8cm}}		
		\centerline{\vspace{0.2cm}\hspace{0.6cm}(e)\hspace{8cm}(f)}
		\centerline{\vspace{-0.2cm}}
		\caption{The figures on the left show the parameter space in agreement with the relic density and the figures on the right show the parameter space in agreement with the relic density and direct detection. In (a) and (b), the $g_v=0.02$ is considered, in (c) and (d), the  $g_v=0.6$, and in (e) and (f), the  $g_v=2$ is considered.} \label{allowed}
	\end{center}
\end{figure}

In this section we show the permissible parameter space in agreement with the given constraints. The range of independent model parameters for scanning
 the results is as follows:
 \begin{itemize}
 	\item The DM mass $M_\psi$ is between $1-2000~\rm GeV$;
    \item The  $M_V$ is between $1-2000~\rm GeV$;
 	\item The mediator scalar mass(dark Higgs) $M_{H_2}$ is between $1-1000~\rm GeV$;
 \end{itemize}
 We also consider three coupling values of 0.02, 0.6 and 2 for $g_v$.
 The first constraint for the parameter space under consideration is the value of the relic density of DM, which must be in agreement with the current observed value by the PLANCK collaboration\cite{Planck:2018vyg}. The value of $\Omega_{DM}$($\Omega_{\psi}$), as reported by the PLANCK collaboration,
carries an estimated theoretical uncertainty around 10 percent, $0.11 \leq \Omega_{DM} h^2 \leq 0.13$. We calculate the relic density numerically by implementing the
model into micrOMEGAs\cite{Alguero:2023zol}. The second constraint is related to direct detection experiments, where we use the results of the XENONnT and LUX-ZEPPELIN(LZ) experiments\cite{XENON:2023cxc,LZ:2024zvo}. The LZ experiment places the strongest possible constraint on direct detection results. The neutrino floor, which is an irreducible background resulting from the scattering of SM neutrinos by nuclei, is also considered\cite{Billard:2013qya}. Finally, we use the ATLAS experimental results that put an upper limit on the invisible Higgs decay to constrain the parameter space\cite{ATLAS:2022yvh}. Figure \ref{allowed}(a) shows the parameter space in agreement with the Relic density.
 Due to the constraint of invisible Higgs decay, the $M_V$ must be greater than  $26~\rm GeV$. Figure \ref{allowed}(b) also shows the allowable points in agreement with the relic density and direct detection. In both of these diagrams, $g_v$ is assumed to be equal to 0.02.
 In Figure \ref{allowed}(c), the permissible parameter space in agreement with the relic density is plotted, and in Figure \ref{allowed}(d), the acceptable points in agreement with the relic density and direct detection for $g_v= 0.6$ are plotted. A similar analysis is plotted in Figures \ref{allowed}(e) and  \ref{allowed}(f) for $g_v =2$.
 Of course, all diagrams include the constraint of invisible Higgs decay. As can be seen, the model is in agreement with the phenomenological constraints for a wide range of parameter space.

\section{ Renormalization group equations and vacuum stability}\label{RGEs}

\begin{figure}
	\begin{center}
		\centerline{\hspace{0cm}\epsfig{figure=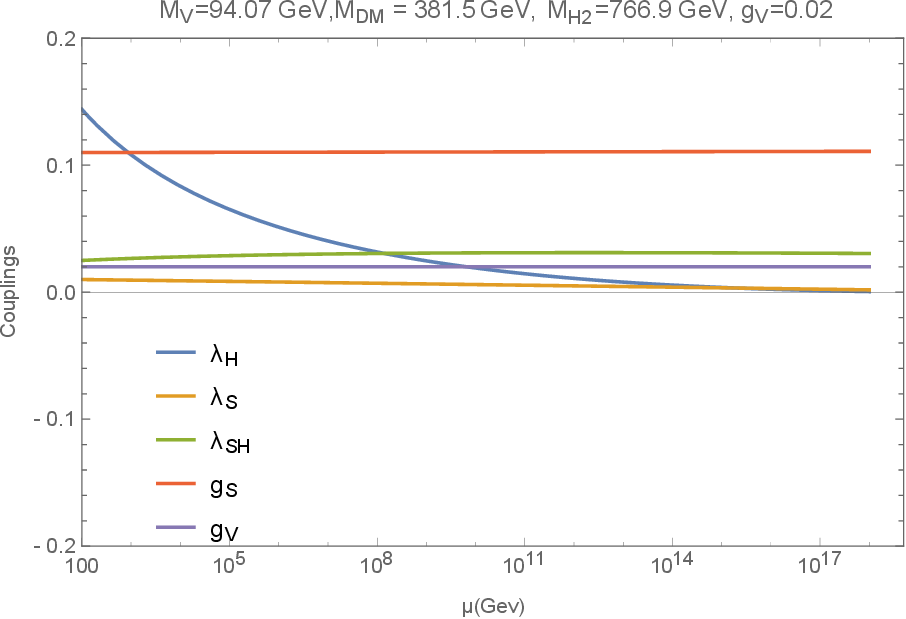,width=13cm}}
        \centerline{\vspace{-0.2cm}}
		\caption{Running of couplings of the model up to Planck scale for Benchmark point in Table \ref{tablerge}.} \label{rge1}
	\end{center}
\end{figure}

In this work, we would like to study the running of the coupling constants with energy.
The $\beta$ functions for our model couplings at one loop are as follow(The SARAH package has been used\cite{Staub:2015kfa}.)
\begin{align}
& (16\pi^2)\beta_{\lambda_{S}} = 20\lambda_{S}^2 +4g_{s,1}^2 \lambda_{S}+4g_{s,2}^2 \lambda_{S} -12\lambda_{S}g_v^2 -2g_{s,1}^4 -2g_{s,2}^4+2\lambda_{SH}^2+6g_v^4  ,\nonumber \\
& (16\pi^2)\beta_{\lambda_{SH}}= -\frac{9}{10}g_{1}^2 \lambda_{SH} -\frac{45}{10}g_{2}^2 \lambda_{SH} +12\lambda_{SH}\lambda_{H}+8\lambda_{SH}\lambda_{S} -4\lambda_{SH}^2 + 2\lambda_{SH}g_{s,1}^2 + 2\lambda_{SH}g_{s,2}^2+6 \lambda_{SH}\lambda_{t}^2  \nonumber ,\\
& (16\pi^2)\beta_{\lambda_{H}}= +\frac{27}{200}g_{1}^4 +\frac{9}{20}g_{1}^2 g_{2}^2 +\frac{9}{8}g_{2}^4 -\frac{9}{5} g_{1}^2 \lambda_{H} -9g_{2}^2 \lambda_{H} +24\lambda_{H}^2 +\lambda_{SH}^2 +12\lambda_{H}\lambda_{t}^2 -6\lambda_{t}^4, \nonumber \\
& (16\pi^2)\beta_{g_{s,1}}=\frac{-3}{2}g_{s,1}g_v^2 +g_{s,1}g_{s,2}^2+2g_{s,1}^3, \nonumber \\
& (16\pi^2)\beta_{g_{s,2}}=\frac{-3}{2}g_{s,2}g_v^2 +g_{s,2}g_{s,1}^2+2g_{s,2}^3, \nonumber \\
& (16\pi^2)\beta_{g_v}= g_v^3.
\label{RGE}
\end{align}
where $\beta_a\equiv \mu \frac{da}{d\mu}$ that $\mu$ is the renormalization scale with initial value $\mu_0=100~\rm GeV$.
The $\beta$ functions of couplings, $g_1$, $g_2$ and $g_3$ are given to one-loop order by:
\begin{align}
& (16\pi^2)\beta_{g_{1}}= \frac{41}{6} g_{1}^3 , \nonumber \\
& (16\pi^2)\beta_{g_{2}}= -\frac{19}{6} g_{2}^3 , \nonumber \\
& (16\pi^2)\beta_{g_{3}}= -7 g_{3}^3 .
\end{align}
Among the Yukawa couplings of SM, the top quark has the largest contribution compared with other fermions in the SM. Therefore, we set all the SM Yukawa couplings equal to zero and consider only the top quark coupling.  The RGE of top quark Yukawa coupling is given to one-loop order by
\begin{align}
(16\pi^2)\beta_{\lambda_t}= -\frac{17}{12}g_{1}^2 \lambda_t -\frac{9}{4}g_2^2 \lambda_{t} -8g_3^2 \lambda_t +\frac{9}{2}\lambda_t^3 .
\end{align}
We choose one representative benchmark which is given in Table \ref{tablerge} and is in agreement with all the experimental constraints presented.
Using the obtained RGEs, the evolution of couplings with energy can be obtained.
Figures \ref{rge1} show the evolution of these couplings. As can be seen, all theoretical constraints in equations \ref{2-16} and \ref{2-17} are verified up to the Planck scale and the stability of the model vacuum is fully guaranteed.

\begin{table}[h]
\centering 
\begin{tabular}{l c c rrrrrrr} 
\hline\hline
 $\#$ &$M_V (GeV)$ &$M_{\psi}(GeV)$ &$M_{H_2}(GeV)$&$g_v$&$g_s$&$\Omega_{DM}h^2$&$\sigma_{DM-N} (cm^2)$&$\sin \alpha$\\
\hline
1&94.07&381.5&766.9&0.02&0.11&0.114&$9.83\times10^{-48}$&0.05 \\
\hline
\end{tabular}
\caption{\label{tablerge}The benchmark point with DM parameters.} 
\end{table}

\section{Veltman condition}\label{Veltman condition}

\begin{figure}
	\begin{center}
		\centerline{\hspace{0cm}\epsfig{figure=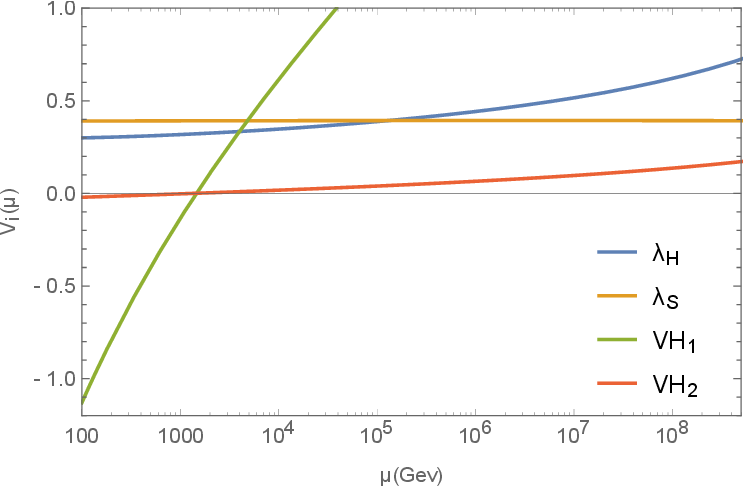,width=11cm}}
        \centerline{\vspace{-0.2cm}}
		\caption{The running of the scalar couplings and the Veltman parameters are shown for the Benchmark of table\ref{tableveltman} in the
parameter space which fulfills the constraints from observed relic abundance, direct detection and theoretical bounds.
The Veltman parameters are vanishing at 1 TeV.} \label{veltmanplot}
	\end{center}
\end{figure}

In order to apply the Veltman condition, we need to obtain the one-loop quadratic mass corrections created in the model for the scalars. For small mixing angles, these corrections are as follows
\begin{align}
&\delta M^2_{H_{1}}= \frac{\Lambda^2}{16\pi^2}[6\lambda_H + \frac{3}{4}g_{1}^2 + \frac{9}{4}g_{2}^2 -6\lambda_{t}^2 + \lambda_{SH}], \nonumber \\
&\delta M^2_{H_{2}}= \frac{\Lambda^2}{32\pi^2}[2\lambda_{SH}+\lambda_{S}+3g_v^2-8g_s^2]
\end{align}
where the Veltmann parameters are:
\begin{align}
& V_{H_{1}}= 6\lambda_H + \frac{3}{4}g_{1}^2 + \frac{9}{4}g_{2}^2 -6\lambda_{t}^2 + \lambda_{SH}, \nonumber \\
& V_{H_{2}}= 2\lambda_{SH}+\lambda_{S}+3g_v^2-8g_s^2 .
\end{align}
Also, because the goal in Veltman approach is to eliminate the strongest divergences (the quadratic divergences) it is sufficient to use the cut-off method \cite{Masina:2013wja,Chakraborty:2012rb}. The corresponding Veltman conditions at energy scale $\Lambda$, are $V_{H_{1}}\sim0$ and $V_{H_{2}}\sim0$ .
we consider one representative benchmark in the viable parameter space in the
our model which respects all the constraints discussed earlier in the paper. This benchmark point is given in table \ref{tableveltman}.
As can be seen from the plots in Fig. \ref{veltmanplot}, the Veltman conditions are satisfied at the scale $\sim$ 1 TeV.
Following the argumentation in \cite{Kolda:2000wi,Barbieri:1987fn}, the measure of fine-tuning as a function of $\Lambda$ is given by
\begin{equation}
\mathcal{F}\equiv \frac{|\delta m^2_{H_1}|}{ m^2_{H_1}}=\frac{\Lambda^2}{16\pi^2 m^2_{H_1}}|V_{H_1}(\Lambda)|.
\end{equation}
Then, we would say that the electroweak scale is fine-tuned to one part in $\mathcal{F}$,
and $\mathcal{F} \leq$ 1 indicates the absence of tuning. For the benchmark point provided above we find $\mathcal{F} \sim$ 0.01 at $\Lambda \sim$ 173 GeV, which indicates no tuning at the electroweak scale.

Of course, it is important to note that what distinguishes the Veltman's approach in our new model from the results of previously presented models is the new parameter space obtained by simultaneously examining the three issues of DM, vacuum stability, and the hierarchy problem. In fact, adding new symmetries or fields to the SM leads to different phenomenology for DM, different potentials for the model, or leads to different RGEs, which differentiates its result from other models.

\begin{table}[h]
\centering 
\begin{tabular}{l c c rrrrrrr} 
\hline\hline
 $\#$ &$M_V (GeV)$ &$M_{\psi}(GeV)$ &$M_{H_2}(GeV)$&$g_v$&$\Omega_{DM}h^2$&$\sigma_{DM-N} (cm^2)$&$\sin \alpha$\\
\hline
1&1507&1145&2325&0.27&0.13&$4.51\times10^{-47}$&0.04 \\
\hline
\end{tabular}
\caption{\label{tableveltman}The benchmark point with DM parameters which satisfy the Veltman conditions.} 
\end{table}

\section{Conclusion}\label{Conclusion}
We have considered an extension of the SM with four new fields: two fermions, a vector and a scalar.
 The fermionic particle constitute our DM candidate, and the scalar particle is considered to be the intermediary between the SM and dark parts.
In this article, we examined three issues: vacuum stability, the hierarchy problem, and DM.
We have found regions in the parameter space which explains correctly the observed DM relic density and is in consistent with the latest bands from direct detection experiments and the invisible Higgs decay. we investigated the vacuum stability and the hierarchy problem within the Veltman approach.
 Upon running of the couplings by solving the RGEs, we consider the two benchmark points in the parameter space which respect both the
vacuum stability and the perturbativity conditions. Another theoretical restriction is to find points which render the Veltman conditions satisfied at energies much
lower than the Planck scale. We found a benchmark point that was in consistent with all the theoretical and experimental constraints presented in the paper.
 we have shown that for this benchmark, the Veltman parameters for all scalars in the model including the Higgs are
vanishing at 1 TeV. The fine-tuning parameter for this benchmark is shown to be much smaller than one, resulting in models with no fine-tuning problem.
We conclude that our model by adding new particles can resolve different
fundamental problems in the SM including the perturbativity and the vacuum stability problem in the SM, the DM problem, and the fine-tuning problem via the Veltman approach.

\bibliography{References}
\bibliographystyle{JHEP}
\end{document}